\documentclass[aps,prl,twocolumn,
showpacs,preprintnumbers,amsmath,amssym,superscriptaddress]{revtex4}
\usepackage{graphicx}
\usepackage{bm}

\begin{document} 

\title{Non-perturbative fixed point in a non-equilibrium phase transition}

\author{L\'eonie Canet}
\affiliation{School of Physics and Astronomy, University of Manchester, Manchester M13 9PL, United Kingdom}

\author{Hugues Chat\'e}
\affiliation{CEA -- Service de Physique de l'\'Etat Condens\'e,~CEN~Saclay,~91191
~Gif-sur-Yvette,~France}

\author{Bertrand Delamotte}
\affiliation{Laboratoire de Physique Th\'eorique de la Mati\`ere Condens\'ee, Universit\'e Paris VI, 75252 Paris Cedex 05, France}

\author{Ivan Dornic}
\affiliation{CEA -- Service de Physique de l'\'Etat Condens\'e,~CEN~Saclay,~91191
~Gif-sur-Yvette,~France}

\author{Miguel A. Mu\~noz}
\affiliation{Instituto~de~F\'\i sica~Te\'orica~y~Computacional~Carlos~I,
~Facultad~de~Ciencias,~Universidad~de~Granada,~18071~Granada,~Spain}

\date{\today}

\begin{abstract}
We apply the non-perturbative renormalization group method to a class
of out-of-equilibrium phase transitions (usually called ``parity
conserving'' or, more properly, ``generalized voter'' class) which is
out of the reach of perturbative approaches. We show the existence of
a genuinely non-perturbative fixed point, i.e. a critical point which
does not seem to be Gaussian in any dimension.
\end{abstract}

\maketitle

Our understanding of equilibrium phase transitions is largely due to
the success of perturbative renormalization group (RG) methods
performed around some (upper or lower) critical dimension $d_{\rm c}$
and to the existence of integrability and conformal symmetry
properties in $d=2$ \cite{Zinn}. The situation is far less
satisfactory out-of-equilibrium, where the relevant ingredients
determining universality classes are sometimes not even known
\cite{Review_abs,Review_PCPD}.  There are a number of technical
reasons for this: (i) many systems possess neither a lower $d_{\rm c}$
nor a low-dimensional exact solution; (ii) even at the critical point,
models with a Langevin-like dynamics, that is, those that involve only
one time derivative in their kinetic terms cannot be conformal
invariant; (iii) contrary to equilibrium, no RG calculation is in
general available at and above three loop order and this prevents from
using series re-summation techniques to compute accurately universal
quantities in low dimensions.  If, moreover, one keeps in mind that
features not accessible to perturbative RG methods may play a crucial r\^ole,
then the so-called non-perturbative renormalization group (NPRG)
approach appears as a method of choice out-of-equilibrium.

Such non-perturbative effects were evidenced recently in a study of
the classic reaction-diffusion problem where particles $A$ diffuse,
branch ($A\to 2A$) and annihilate ($2A\to\emptyset$) with rates $D$,
$\sigma$, and $\lambda$ \cite{BARW_prl}. Whereas, in an important work
\cite{tauber}, Cardy and T\"auber had shown that perturbative RG
calculations led to conclude that no finite-$\sigma$ transition to the
empty absorbing state is possible for $d>2$, an NPRG study at the
non-universal level showed that such absorbing phase transitions exist
in any finite dimension (with their critical properties in the
directed percolation class, as expected from perturbative RG).

In this Letter, we apply the NPRG method to a similar class of
absorbing phase transitions, one for which even the {\it universal}
properties are out of the reach of perturbative approaches. We put
forward the existence of a genuinely non-perturbative fixed point,
i.e. a critical point which does not seem to be Gaussian in any
dimension. Our calculations unveil the structure of the RG flow,
reveal clearly why perturbative methods are doomed to failure in this
case, and provide estimates of critical exponents heretofore
accessible only via numerical simulations or series expansions.

We actually treat two classes of systems known to be equivalent in
$d=1$, the physical dimension where the non-perturbative fixed point
alluded to above is relevant. The first group \cite{PC,PC_models}
includes the reaction-diffusion system described above but where the
branching reaction now creates pairs of particles ($A\to 3A$), so
that, incidently, the parity of the total number of particles is
conserved. Improperly named ``parity-conserving'' (PC) class (as
argued in \cite{Kockel1}, where this conservation law was shown to
have no influence on similar reaction-diffusion systems), it is best
characterized by the second group of problems \cite{GV_models}, that
of phase transitions into one out of two $Z_2$-symmetric absorbing
states \cite{Misc}.  In $d=1$, the particles of the PC model can
indeed be seen as interfaces between ``$+$'' and ``$-$'' domains
\cite{drofeli}. 
In this ``spin'' language, domains evolve and compete under 
$Z_2$-symmetric rules with noise only acting at interfaces, 
 the definition of the ``generalized voter'' (GV) class 
as given in \cite{Voter-prl}.

The reaction-diffusion problem ($A\to 3A$, $2A\to\emptyset$) was also
studied in \cite{tauber}.  Using the Doi-Peliti formalism \cite{doipeliti}, one
can obtain the
following microscopic action  \cite{PC,tauber}:
\begin{equation}
S[\phi,\bar\phi]= \int_{x,t}\! \bar\phi (\partial_t\! -\! D
\nabla^2)\phi - \lambda (1\!-\!\bar\phi^2)\phi^2 + \sigma
(1\!-\!\bar\phi^2)\phi \bar\phi
\label{bareaction}
\end{equation}
in terms of the ``physical'' density field $\phi$ and the associated
response field $\bar\phi$.  Cardy and T\"auber first performed an expansion
around the upper critical dimension $d_{\rm c}=2$ where the transition
occurs at zero branching rate $\sigma$, so that the fixed point then
is that of the pure annihilation problem $2A\to\emptyset$.  They
showed that this annihilation fixed point remains relevant down to
$d=\frac{4}{3}$ (at one-loop order), where it becomes also attractive
in the $\sigma$ direction. In another  expansion,
this time performed directly in $d=1$, they were able to identify an 
appropriate combination of the coupling constants $\lambda, \sigma$ which 
does admit a fixed point for $d \leq
\frac{4}{3}$,
although the flow diagram with respect to these original variables
is rather peculiar. Even if their results do suggest that 
a phase transition should exist at $\sigma \neq 0$ 
---as observed in numerical simulations and mean-field-based expansions of
many microscopic models in $d=1$ \cite{PC_models,GV_models}---,
the critical
exponents remain poorly determined and, worse, the very possibility of
computations beyond one loop-order appears to be problematic \cite{tauber}.

As for the GV class, the following Langevin equation was recently
proposed \cite{drofeli}:
\begin{equation}
\partial_t\phi= (-\sigma \phi + \mu  \phi^3) (1- \phi^2)+ D\nabla^2\phi +
 \sqrt{2\lambda(1-\phi^2)}\,\eta
\label{eq-drofeli}
\end{equation} 
with $\phi \in [-1,1]$ and $\eta$ is a delta-correlated Gaussian noise
of unit variance. The $(1-\phi^2)$ factors, appearing both in the
deterministic force and in the noise amplitude, impose $\phi=\pm~1$ to
be symmetric absorbing states.  In $d=1$, only one type of transition
is observed by varying $\sigma$ for any value of $\mu$, and its
critical properties are indeed those observed for the so-called
parity-conserving models \cite{NOTE3}.  Taking $\mu=0$ for the sake of
simplicity, the generating functional associated with the simplified
Langevin equation is nothing but (\ref{bareaction}) where $\phi
\leftrightarrow \bar{\phi}$ and $t\to -t$. Thus, at this bare level,
the two problems are strictly equivalent in $d=1$. We shall see later
that this equivalence is preserved under the RG flow.

We cannot detail here the implementation of the NPRG but only mention
the essential features of the method \cite{Berges02,bagnuls01}. The main idea is
to build a one-parameter family of models, indexed by a momentum scale
$k$, interpolating smoothly between the short-distance physics at the
(microscopic) scale $k=\Lambda$, where no fluctuation has been taken
into account, and the long-distance physics at scale $k=0$, where all
fluctuations have been integrated out. In Wilson's original
formulation, this leads to a flow of effective Hamiltonians --- for the
slow modes --- defined at scale $k$. Following \cite{Berges02}, we focus on the flow of
``free energies'' for the rapid modes, $\vert q\vert \in [k,\Lambda]$,
i.e. those already integrated out at this scale. This is achieved by
adding a mass-like term of order $k^2$ to the slow modes ($\vert
q\vert < k$), which ``freezes'' 
 them.  This mass term reads:
\begin{equation}
\Delta S_k[\phi,\bar\phi] =\int_{q,\omega} R_k(q^2)\bar\phi(-q,-\omega)
 \phi(q,\omega)
\end{equation}
where a convenient choice \cite{litim} of ``cut-off'' function is
$$R_k(q^2)=k^2(1-q^2/k^2) \theta(1-q^2/k^2) \;.$$ 
The ``partition
functions'' $Z_k[J,\bar{J}]=\int D\phi D\bar\phi
\exp(-S-\Delta S_k +\int J\phi +\int \bar{J}\bar\phi)$ become therefore
$k$-dependent. Through the Legendre transform of   
$\log Z_k[J,\bar{J}]$, one obtains the state function $\Gamma_k$ --- analogous to the
Gibbs free energy at equilibrium --- which depends on the fields
$\psi=\delta \log Z_k/\delta J$ and $\bar\psi=\delta \log Z_k/\delta
\bar J$:
\begin{equation}
\Gamma_k[\psi,\bar\psi] +\log Z_k[J,\bar{J}]= 
\int J \psi +\int \bar{J} \bar{\psi} -\int R_k \psi \bar\psi.
\label{gammak}
\end{equation}
Note that the last term in Eq.(\ref{gammak}), proportional to $R_k$, ensures
that $\Gamma_k$ has the proper limit at $k=\Lambda$: $\Gamma_{k=\Lambda}\sim
S$ \cite{Berges02}.  The following exact functional differential equation governs
the RG flow of $\Gamma_k$ under an infinitesimal change of
the scale $s=\log(k/\Lambda)$
\cite{Berges02,canet}:
\begin{equation}
\partial_s \Gamma_k = \frac{1}{2} {\rm Tr} \int_{q,\omega} \partial_s \hat{R}_k \left(\hat\Gamma_k^{(2)} + \hat{R}_k\right)^{-1},
\label{equationrg}
\end{equation}
where $\hat{R}_k$ is the symmetric, off-diagonal, $2\times 2$ matrix
of element $R_k$ and $\hat\Gamma_k^{(2)}[\psi,\bar\psi]$ the $2\times
2$ matrix of second derivatives of $\Gamma_k$ w.r.t. $\psi$ and
$\bar\psi$. Obviously, Eq.~(\ref{equationrg}) cannot be solved exactly
and one usually truncates it. A standard truncation is the derivative
expansion \cite{Berges02} in which $\Gamma_k$ is expanded as a power
series in $\nabla$ and $\partial_t$.  The local potential
approximation (LPA), which is the simplest such truncation, consists
in keeping only a potential term in $\Gamma_k$ while neglecting any
field renormalization:
\begin{equation}
\Gamma_k^{\rm LPA}=\int_{x,t} \left\{U_k(\psi,\bar\psi) + 
\bar\psi  (\partial_t - D \nabla^2)\psi \right\}.
\end{equation}
If the anomalous dimensions are not too large, the LPA already
provides a good description of the effective potential as well as a 
rather accurate estimate of the exponent $\nu$ governing the divergence 
of the correlation length. 
Since our main goal is to identify the non-perturbative fixed point
governing the PC/GV transition in $d=1$,  we shall
  restrict ourselves, in what 
follows, to the LPA \cite{LASTNOTE}.
 
The NPRG equation for the effective potential, valid for all reaction-diffusion processes involving a single species, has been established
 in  \cite{canet}.
 Studying a particular model
 amounts to solving this equation in a sub-space defined by the
symmetries of the problem, starting with the corresponding
  microscopic action $S$.
 The flow equation for the dimensionless potential $u=k^{d+2}U_k$,
expressed in terms of the dimensionless fields $\psi\to k^{-d}\psi$
and $\bar\psi\to\bar\psi$, reads
(to  lighten  notations, we omit the implicit dependence of 
$u$ on the running scale):
\begin{equation}
\partial_su = -(d\!+\!2)u + d\,\psi\, u^{(1,0)} -
V_d \left[1-\frac{u^{(2,0)} u^{(0,2)}}{(1\!+\! u^{(1,1)})^2}\right]^{-\frac{1}{2}},
\label{flowpot}
\end{equation}
where  
$u^{(n,p)}=\frac{\displaystyle\partial^{n+p}u}{\displaystyle\partial^n\psi\partial^p\bar\psi}$
and 
$V_d=\frac{\displaystyle 2^{-d+1}\pi^{-d/2}}{\displaystyle d\, \Gamma(d/2)}$.
 In our problem, the effective potential must remain unchanged under the
simultaneous transformations $\psi\to -\psi $ and $\bar\psi\to
-\bar\psi$, (``parity-conservation''/$Z_2$ symmetry of the PC/GV
models) \cite{tauber,NEWNOTE}.  This leads to the existence of three quadratic
invariant quantities, $\psi^2$, $\bar\psi^2$, and $\psi\bar\psi$, from
which all other invariant combinations of the fields can be
built. Action (\ref{bareaction}) can be expressed in terms of these
invariants, but it also possesses additional features: the potential
of the microscopic action $S$ is proportional to $1-\bar\psi^2$ and
vanishes for $\psi=0$ (in the PC language). One can check that this
structure is preserved by the renormalization flow, which further
constrains the functional subspace in which the running potential
evolves.  To summarize, the structure of the running potential
defining our problem is
\begin{equation}
\label{pde_gvpc}
u(\psi,\bar\psi)=(1-\bar\psi^2){\cal F}(\psi^2,\psi\bar\psi) \;\;{\rm
with}\;\; {\cal F}(0,0)=0 \;.
\end{equation}

Postponing the numerical resolution of the partial differential
equation (\ref{flowpot}) in the functional subspace defined above, we
now perform a Taylor series expansion of the potential.  In the
absence of any information about the radius of convergence, the point
around which the expansion is performed matters, all the more so since
we want to eventually truncate it.
Here, we shall expand the potential  around a non-negative
solution of the
stationary equations of motion $\frac{\partial u}{\partial \psi}=0$
and $\frac{\partial u}{\partial \bar \psi}=0$. The former 
 is always satisfied
by $\bar\psi=1$, while the  
solutions of the latter  then correspond
 to either  $\psi=0$ (the ``origin'') or to a (running) $\psi > 0$
(``the minimum'').
  We present results obtained around the origin,
which involve lighter  equations than those obtained around the
minimum.

\begin{figure}
\includegraphics[width=76mm,clip]{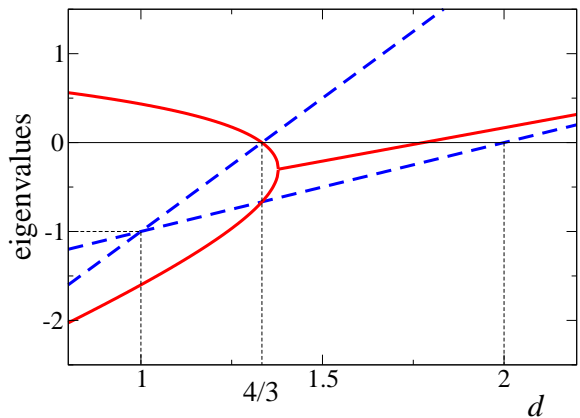}
\caption{(Color online)
Variation with $d$ of the eigenvalues 
of the fixed points in the lowest-order LPA. Blue, 
dashed lines: pure annihilation
fixed point $F_{\rm A}$. Red, solid lines: non-perturbative fixed point $F^*$  
whose eigenvalues are both negative for $\frac{4}{3}<d<1.3784\ldots$ 
and complex-conjugated at larger $d$ (only the real part is plotted).}
\label{fig-vp}
\end{figure}

\begin{figure}
\includegraphics[width=76mm,clip]{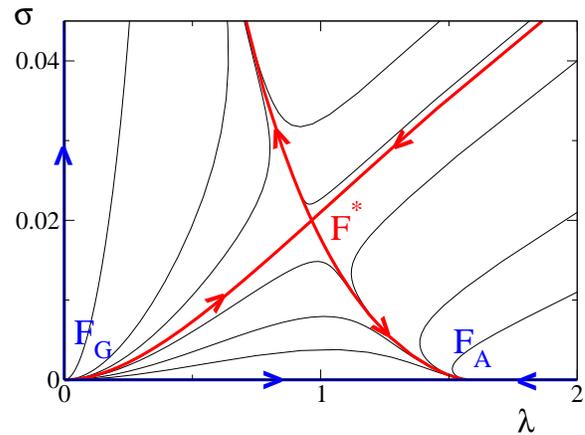}
\caption{(Color online)
Flow diagram of the lowest-order LPA in $d=1$ (as usual, arrows
represent the RG evolution as $s$ is decreased towards the ``infra-red'',
macroscopic
 limit $k \to 0$).}
\label{fig-flot}
\end{figure}

The simplest truncation, that we now analyze in some detail, 
consists in keeping in $u$ only the two coupling constants
already present in $S$. 
Inserting this Ansatz in Eq.(\ref{flowpot}), we obtain the following 
non-trivial flows for the running constants\cite{NOTE}:
\begin{eqnarray}
\partial_s \lambda& =& -\lambda\,(2-d)+ 2 V_d\,\frac{\lambda^2(1+22\sigma)}{(1-2\sigma)^3}\\
\partial_s \sigma& =& -2\,\sigma+ 6 V_d\,\frac{\lambda\sigma}{(1-2\sigma)^2} \;.
\label{flow}
\end{eqnarray}
This RG flow possesses three fixed points:
the trivial, Gaussian, fixed point $F_{\rm G}$ ($\lambda^*_{\rm G}=\sigma^*_{\rm G}=0$),
the annihilation fixed point $F_{\rm A}$ ($\lambda^*_{\rm A}=\frac{2-d}{2V_d},
\sigma^*_{\rm A}=0$), and the non-trivial fixed point $F^*$ of coordinates:
\begin{equation}
\lambda^*=\frac{192}{V_d(28-3d)^2} \,, \;\;\; \sigma^*=\frac{4-3d}{56-6d} \; .
\end{equation}
The Gaussian fixed point $F_{\rm G}$, of eigenvalues $(2,2-d)$ is
relevant above $d_{\rm c}=2$, where it coincides with $F_{\rm A}$. For
$d\in[\frac{4}{3},2]$, $F_{\rm A}$, whose eigenvalues are
$(d-2,3d-4)$, is relevant. At $d=\frac{4}{3}$, $F_{\rm A}$ and the
non-trivial fixed point $F^*$ coincide and exchange stability, so that
$F^*$ is the relevant fixed point for $d<\frac{4}{3}$
(Fig.~\ref{fig-vp}). Note that then $\sigma^*>0$, and thus $F^*$ is in
the physical region of parameter space, whereas it plays no role for
the physics of reaction-diffusion systems when $d>\frac{4}{3}$. Note
also that $F^*$ is not Gaussian in any dimension (at least at this
order), and is thus out of the reach of any perturbative expansion.
The PC/GV fixed point in $d=1$ is thus $F^*$, a genuinely
non-perturbative fixed point. Its associated critical exponent, given
by the inverse of its negative eigenvalue is
$\nu=\frac{12}{\sqrt{149}-7}\simeq 2.30$. The flow diagram in this
dimension is shown in Fig.~\ref{fig-flot}. The once unstable manifold of
$F^*$, connected to $F_{\rm G}$, is the critical ``surface''
separating the absorbing and the active phases.  The flow around
$F_{\rm A}$ is rather peculiar: as $d$ is decreased from
$\frac{4}{3}$, the eigenvector of $F_{\rm A}$ which is not parallel to
the $\lambda$ axis rotates and becomes parallel, in $d=1$, to this
axis. Thus, $F_{\rm A}$ is degenerate for $d=1$, since its two
eigenvectors coincide (Fig.~\ref{fig-vp})\cite{NOTE2}.  This implies
in particular that every point flowing in the absorbing phase reaches
$F_{\rm A}$ along the $\lambda$ axis. It is not clear to us, at this
point, what might be the physical signature of this for microscopic
models.

\begin{table}[htbp]
\caption{Values of exponent $\nu$ with the order $n$ of the LPA 
truncation (see text). The column ``min'' refers to the minimal
truncation $(\lambda,\sigma)$. The last column is a conservative estimate
taken from various Monte-Carlo simulations.}
\label{nu}
\begin{ruledtabular}
\begin{tabular}{cccccccccc}
$n$ & min & 3 & 4 & 5 & 6 & 7 & 8 & 9 & MC \\
\hline
$\nu$   & 2.30 & 2.48 & 2.20 & 2.23 & 2.11 & 2.057 & 2.015 & 2.0017 & 1.85(10)\\
\end{tabular}
\end{ruledtabular}
\end{table}

We now report on the results obtained for truncations of the potential $u$
that go far beyond the simplest truncation described above.
Of course, there exist many  ways to organize a 
polynomial expansion of $u$  around $\psi=0, \bar\psi^2=1$
in terms of the three quadratic invariants
$\psi \bar \psi$, $1- \bar \psi^2$, $\psi^2$
 which abides the $Z_2$-symmetry of the PC/GV class.
Equivalently, thanks to Eq.(\ref{pde_gvpc}), one can use any basis 
spanned by monomials in
$\psi^2,\psi \bar \psi$, and we have tried several choices for it.
In all cases,  we stress that the qualitative picture
unveiled at the minimal level is preserved at higher orders.
It turns out that the fastest convergence of the exponent $\nu$
is obtained with the basis which is also the more transparent 
from the physical point of view. At order
``$n$'', we consider all  
 possible branching  $(n-2 m)A \to n A$ and annihilating reactions 
$n A \to (n- 2 k)A$ involving  at most $n$ particles. These
elementary reactions respectively correspond to  all the terms
$(1-{\bar \psi}^{2 m}){(\psi \bar \psi)}^{n-2m}$ and
$(1-{\bar\psi}^{2 k}) \psi^n {\bar \psi}^{n-2k}$,
 and are anyhow ineluctably generated under renormalization.
Table~\ref{nu} shows the exponent $\nu$ computed up to order 9. 
The convergence is rather good and we deduce
that within the LPA $\nu= 2 \pm 0.1$, already in fair agreement with the 
results obtained in numerical simulations.
We finally report a rather surprising fact: 
at all orders of the truncation, the dimension at which
$F^*$ and $F_{\rm A}$ cross is $d=\frac{4}{3}$. 
On the other hand we know, from the perturbation expansion 
performed around $d=2$ that this dimension is no longer $\frac{4}{3}$ 
at two-loop order so that there is no reason to believe 
that this value is an exact result. Thus the deviation from 
$\frac{4}{3}$ can only come from orders beyond the LPA
in the derivative expansion.

To summarize, we have shown, within the local potential approximation of the 
NPRG, that the critical point of the PC/GV class
of absorbing phase transitions is a genuinely non-perturbative fixed point,
out of reach of perturbative methods.
This result will have to be 
refined within higher-orders approximation of the derivative expansion,
so that a full set of critical indices can be estimated, 
not just the correlation length exponent $\nu$.
Finally, ongoing work aims at a full numerical simulation of 
Eq.~(\ref{flowpot}) (for the current problem and similar ones, such 
as the directed percolation class), which should allow to fully confirm these
results and also offer access to other currently much
debated reaction-diffusion
problems, such as the pair contact process with diffusion \cite{Review_PCPD}.

We thank  Uwe~C.~Ta\"uber for his keen interest on our work.
The Institut Carlos I and the University
of Granada have  provided generous support and hospitality 
when part of this work was completed.
M.~A.~M. also acknowledges financial support from the Spanish MCyT (FEDER)
under project BFM2001-2841.


\begin{thebibliography}{99}
\bibitem{NEWNOTE} Note that discarding the (somewhat trivial
and arbitrary) dimensional part of Eq.~(\ref{flowpot}),
 the flow equation of the effective potential
is invariant under the exchange $\psi \leftrightarrow \bar\psi$. 
The equivalence 
between the PC and GV classes is thus preserved under renormalization.
\bibitem{NOTE} Eqs.~(\ref{flow}) show a pole at $\sigma=\frac{1}{2}$. 
However, if we integrate the RG flow from a region of small $\sigma$,
it never hits the pole. We also find that the relevant fixed point
always lies in the small-$\sigma$ region. 
In fact, the pole would be problematic only if we wanted
to follow the flow of a microscopic action 
with $\sigma\geq \frac{1}{2}$ from $k=\Lambda$.
In this case, our equations would have to be modified 
in the large $k$ regime since then a cut-off function $R_k$ diverging 
at $k=\Lambda$ would have to be taken, 
which would automatically eliminate the pole. Note that at any rate,
such modifications do not change the fixed points, exponents, etc.
\bibitem{NOTE2} Cardy and Ta\"uber found from their RG analysis  \protect\cite{tauber}
of the ``massive theory'' performed at fixed 
dimension a flow in $d=1$ which differs qualitatively from ours, 
albeit  the two eigendirections  around 
$F_{\rm A}$ were also  parallel there.
\end{thebibliography}
\end{document}